\documentclass[preprint]{elsarticle}

\usepackage{graphicx}
\usepackage{amssymb}
\usepackage{amsthm}
\usepackage{amsmath}
\usepackage{color}

\journal{Nuclear Instruments and Methods in Physics Research Section A}

\biboptions{sort&compress}

\begin{document}

\begin{frontmatter}

\title{Separated flow operation of the SHARAQ spectrometer
       for in-flight proton decay experiments}


\author[cns]{M.~Dozono\corref{cor1}} \ead{dozono@cns.s.u-tokyo.ac.jp}
\author[rnc]{T.~Uesaka}
\author[cns]{S.~Michimasa}
\author[cns]{M.~Takaki}
\author[cns]{M.~Kobayashi}
\author[cns]{M.~Matsushita}
\author[cns]{S.~Ota}
\author[cns]{H.~Tokieda}
\author[cns]{S.~Shimoura}

\address[cns]{Center for Nuclear Study, University of Tokyo, Saitama 351-0198, Japan}
\address[rnc]{RIKEN Nishina Center, Saitama 351-0198, Japan}

\cortext[cor1]{Corresponding author}

\begin{abstract}
  New operation mode, ``{\it separated flow mode}'', has been developed 
  for in-flight proton decay experiments with the SHARAQ spectrometer. 
  In the separated flow mode,
  the protons and the heavy-ion products 
  are separated and measured in coincidence 
  at two different focal planes of SHARAQ. 
  The ion-optical properties of the new mode were studied
  by using a proton beam at 246~MeV, 
  and the momentum vector was properly reconstructed
  from the parameters measured in the focal plane of SHARAQ. 
  In the experiment with the $({}^{16}{\rm O},{}^{16}{\rm F})$ reaction
  at a beam energy of 247~MeV/u, 
  the outgoing ${}^{15}{\rm O}+p$ produced by the decay of ${}^{16}{\rm F}$
  were measured in coincidence with SHARAQ. 
  High energy resolutions of 100~keV (FWHM) and $\sim 2~{\rm MeV}$
  were achieved for the relative energy of 535~keV, 
  and the ${}^{16}{\rm F}$ energy of 3940~MeV, respectively. 
\end{abstract}


\end{frontmatter}


\section{Introduction}

The availability of Radioactive Isotope (RI) beams 
has made it possible to study exotic properties of 
nuclei far from the $\beta$-stability line 
as well as to investigate key nuclear reactions relevant to 
important astrophysical phenomena. 
Among various experimental methods with RI beams, 
the invariant-mass method has been extensively used 
for the spectroscopy of particle-unbound states in exotic nuclei. 
In the method, 
the excitation energy of a given state is obtained 
by reconstructing the invariant mass of all the decay products. 
Thus it is technically essential to
detect heavy-ion fragments and nucleons in coincidence. 
This type of measurement has been performed efficiently 
by the use of spectrometers having large solid-angle and momentum acceptance. 
For example, in Ref.~\cite{PhysRevLett.83.2910}, 
the KaoS spectrometer~\cite{Senger1993393} at GSI was used 
for a coincidence detection of ${}^{7}{\rm Be}$ and a proton 
from the Coulomb dissociation of a ${}^{8}{\rm B}$ beam.
At RIKEN, 
a simple dipole magnet was used in 
combination with the neutron-detector array 
based on plastic scintillators 
to study the unbound states of 
neutron-rich nuclei such as
${}^{11}{\rm Li}$~\cite{PhysRevLett.96.252502}, 
${}^{13}{\rm Be}$~\cite{Kondo2010245}, 
and so on. 
In the new facility at RIKEN, 
RI Beam Factory (RIBF), 
the advanced spectrometer setup, 
SAMURAI~\cite{1742-6596-312-5-052022,Kobayashi2013294} 
has been constructed. 
The SAMURAI spectrometer has been designed to make efficient 
invariant-mass measurements of both neutron- and proton-unbound states.
Some measurements for neutron-unbound states  
have already been performed by detecting 
a few neutrons in coincidence with a heavy-ion fragment~\cite{Kobayashi2013294}. 

Unlike large-acceptance spectrometers such as SAMURAI and KaoS, 
the SHARAQ spectrometer~\cite{Uesaka20084218,Uesaka201203C007,Michimasa2013305} at RIBF 
enables the high-resolution analysis of the reaction products, 
which is helpful for various pricise measurements such as 
the particle identification for heavy isotopes, 
the momentum distribution measurements via knockout reactions, 
the Q-value measurements via multi-nucleon transfer reactions, 
and so on. 
SHARAQ also open up new experimental possibilities 
in combination with the high-resolution beam-line~\cite{Kawabata20084201}.
One interesting example is a new missing mass spectroscopy 
with an RI beam used as a probe~\cite{Uesaka201203C007}. 
Since RI beams have a variety of isospin, spin, and
internal energy (mass excess) values, 
RI-beam induced reactions have unique sensitivities
that are missing in stable-beam induced reactions 
and can be used to reach yet-to-be-discovered
states~\cite{Uesaka01102012,1742-6596-445-1-012018}.
With SHARAQ, 
investigations of spin-isospin properties in nuclei 
have been strongly promoted 
by RI-induced charge-exchange reactions such as 
$(t,{}^{3}{\rm He})$~\cite{PhysRevLett.108.262503}, 
$({}^{12}{\rm N},{}^{12}{\rm C})$~\cite{Noji2012},
$({}^{10}{\rm C},{}^{10}{\rm B})$~\cite{Sasamoto2012}, and 
$({}^{8}{\rm He},{}^{8}{\rm Be})$~\cite{Kisamori2015}.

We have developed a new ion-optical mode of SHARAQ 
for the coincident measurement between proton and heavy-ion fragments. 
The new mode called ``{\it separated flow mode}'' enables 
the invariant-mass spectroscopy of proton-unbound states with SHARAQ, 
and thus extends the research field in the nuclear chart 
toward proton-rich nuclei. 
In addition, proton-unbound nuclei can be used as probe particles.
One interesting example is the parity-transfer reaction
$({}^{16}{\rm O},{}^{16}{\rm F}(0^-,{\rm g.s.}))$~\cite{Dozono2012}.
This reaction has a unique sensitivity to unnatural parity states,
and can be used as a powerful tool 
to probe $0^-$ states in a target nucleus.

In this paper, we will describe a new ion-optical mode,
separated flow mode, applied to SHARAQ
for the coincidence measurement between proton and heavy-ion fragments. 
The ion-optical design of the new mode is outlined 
in Section~\ref{sec:ion_optical_design}.
The experimental reconstruction of the momentum vector of the proton
from the parameters measured in the focal plane of SHARAQ 
is given in Section~\ref{sec:experiments}. 
The performance of the new mode is also described 
by taking the recent experiment on the 
$({}^{16}{\rm O},{}^{16}{\rm F})$ reaction as an example.

\section{Ion-optical design}
\label{sec:ion_optical_design}

\begin{figure}[t]
\centering
\includegraphics[width=0.9\linewidth]{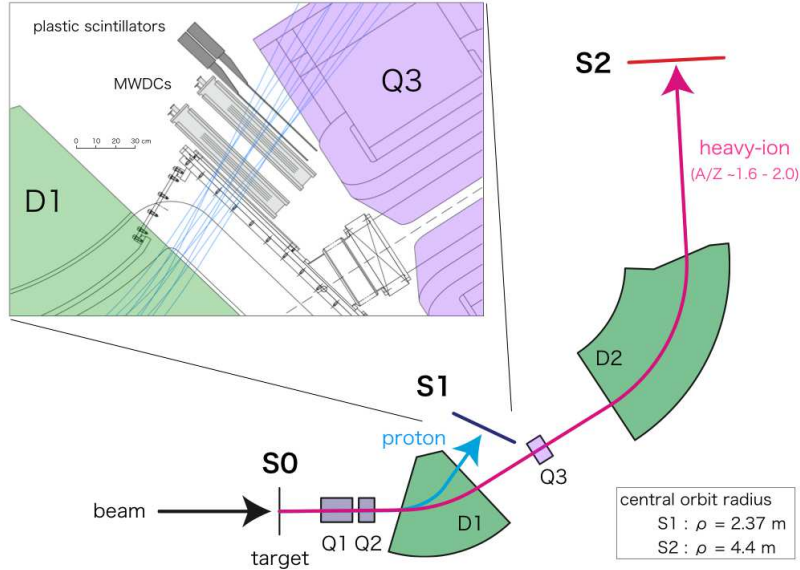}
\caption{\label{fig:setup}
Schematic layout of the SHARAQ spectrometer.
}
\end{figure}

The SHARAQ spectrometer consists of 
two superconducting quadrupole magnets (Q1 and Q2), 
one normal conducting quadrupole magnet (Q3) and 
two dipole magnets (D1 and D2) 
in a ``QQDQD'' configuration (See Fig.~\ref{fig:setup}). 
This spectrometer is designed to 
achieve high momentum and angular resolutions of 
$\delta p/p \sim 1/14700$ and $\theta \sim 1~{\rm mrad}$, respectively. 
Details of the ion-optical and magnet designs can be found in
Refs~\cite{Uesaka20084218,Uesaka201203C007,Michimasa2013305}. 

In ``{\it separated flow mode}'', 
the SHARAQ spectrometer is used 
as two spectrometers with different magnet configurations, 
``Q1-Q2-D1'' and ``Q1-Q2-D1-Q3-D2''. 
The reaction products from the target (S0) 
are separated and analyzed in either of these two configurations 
depending on their $A/Z$ values. 
The particles with $A/Z \sim 1$ such as protons 
are analyzed in the ``Q1-Q2-D1'' configuration 
and detected at the S1 focal plane, 
which is at the low-momentum side of the exit of the D1 magnet. 
On the other hand, 
the heavy-ion products are 
analyzed in the ``Q1-Q2-D1-Q3-D2'' configuration 
to increase the resolving power, 
and detected at the S2 final focal plane. 
Therefore, this new technique enables us 
to perform the coincidence measurements 
of the proton and heavy-ion pairs 
produced from the decays of proton-unbound states in nuclei. 

In this new mode of SHARAQ, 
we have planned exeperiments with 
the parity-transfer 
$({}^{16}{\rm O},{}^{16}{\rm F}(0^-,{\rm g.s.}))$ reaction. 
In these experiments, 
the outgoing ${}^{15}{\rm O}+p$ pairs produced 
in the decay of ${}^{16}{\rm F}$ are measured in coincidence.
For the clear identification of ${}^{16}{\rm F}(0^-)$, 
a high relative energy resolution of 
$\sim 100~{\rm keV}$ at $E_{\rm rel}=535~{\rm keV}$ is required. 
Furthermore, 
a energy resolution of ${}^{16}{\rm F}$ 
should be better than $\sim 2~{\rm MeV}$ 
to use the reaction as 
a missing-mass spectroscopy tool. 
These requirements can be achieved 
with the momentum and angular resolutions of 
$\delta p/p < 1/300$ $(1/4500)$ and 
$\theta < 4 $ $(4)~{\rm mrad}$ for proton (${}^{15}{\rm O}$) 
for a beam energy of $250~{\rm MeV/u}$.  

In order to achieve sufficient momentum and angular resolutions, 
we consider the ion-optical properties of the spectrometer 
in the transfer-matrix formalism. 
The trajectory of a particle from the target to the focal plane 
can be described by using the first-order terms of transfer matrix as 
\begin{equation}
  \label{eq:trajectory}
  \begin{pmatrix}
    x_{\rm fp} \\
    a_{\rm fp} \\
    y_{\rm fp} \\
    b_{\rm fp} \\
    \delta
  \end{pmatrix}
  =
  \begin{pmatrix}
    (x|x) & (x|a) & 0 & 0 & (x|\delta) \\
    (a|x) & (a|a) & 0 & 0 & (a|\delta) \\
    0 & 0 & (y|y) & (y|b) & 0 \\
    0 & 0 & (b|y) & (b|b) & 0 \\
    0 & 0 & 0 & 0 & 1 
  \end{pmatrix}
  \begin{pmatrix}
    x_{\rm tgt} \\
    a_{\rm tgt} \\
    y_{\rm tgt} \\
    b_{\rm tgt} \\
    \delta
  \end{pmatrix}  ,
\end{equation}
where $x_{\rm fp(tgt)}$ and $a_{\rm fp(tgt)}$ 
are the horizontal position and angle of a particle
at the focal plane (target), 
and $y_{\rm fp(tgt)}$ and $b_{\rm fp(tgt)}$
are the vertical position and angle, respectively. 
$\delta=\Delta p/p$ is a fractional momentum deviation
from the central trajectory. 
For the spectroscopy, 
one needs to reconstruct the momentum ($\delta$) 
and the angle at the target position ($a_{\rm tgt}$, $b_{\rm tgt}$)
by solving Eq.~(\ref{eq:trajectory}). 
In the focal plane, the $(x|a)$ term vanishes. 
Furthermore, if the beam spot size at the target position is small enough, 
the contributions from the terms of $(x|x)$, $(a|x)$ and $(y|y)$ are negligible. 
Then Eq.~(\ref{eq:trajectory}) can be easily solved as 
\begin{align}
  \delta     & =  \frac{x_{\rm fp}}{(x|\delta)}, \label{eq:delta_reconst}\\
  a_{\rm tgt} & =  \frac{a_{\rm fp}-(a|\delta)\delta}{(a|a)}, \label{eq:a_reconst}\\
  b_{\rm tgt} & =  \frac{y_{\rm fp}}{(y|b)}. \label{eq:b_reconst}
\end{align}
Therefore, $\delta$, $a_{\rm tgt}$ and $b_{\rm tgt}$ can be reconstructed
from $x_{\rm fp}$, $a_{\rm fp}$ and $y_{\rm fp}$, respectively. 

In Eqs.~(\ref{eq:delta_reconst})--(\ref{eq:b_reconst}), 
we assume that $x_{\rm tgt}=y_{\rm tgt}=0$ 
because of the small beam spot size of $\Delta x_{\rm tgt}$ and $\Delta y_{\rm tgt}$. 
This assumption brings the ambiguities of $\delta$, $a_{\rm tgt}$, and $b_{\rm tgt}$.
The ambiguities 
$\Delta \delta$, $\Delta a_{\rm tgt}$, and $\Delta b_{\rm tgt}$
can be estimated as 
\begin{align}
  \Delta \delta     & = \frac{(x|x) \Delta x_{\rm tgt}}{(x|\delta)}, \label{eq:delta_uncertainty}\\
  \Delta a_{\rm tgt} & = \frac{(a|x) \Delta x_{\rm tgt}}{(a|a)}, \label{eq:a_uncertainty} \\
  \Delta b_{\rm tgt} & = \frac{(y|y) \Delta y_{\rm tgt}}{(y|b)}.\label{eq:b_uncertainty}
\end{align}
$\Delta x_{\rm tgt}$ and $\Delta y_{\rm tgt}$ are typically 1-2~mm. 
Thus, our requirements can be achieved with
the conditions of 
\begin{align}
  (x|\delta)/(x|x) & > 0.6~{\rm m}, \label{eq:s1_condition_1} \\
  \sqrt{\left( \frac{(a|x)}{(a|a)}\right)^2
    +\left( \frac{(y|y)}{(y|b)}\right)^2}  & < 2~{\rm rad/m}, \label{eq:s1_condition_2}
\end{align}
for proton, and 
\begin{align}
  (x|\delta)/(x|x) & > 9.0~{\rm m}, \label{eq:s2_condition_1}\\
  \sqrt{\left( \frac{(a|x)}{(a|a)}\right)^2
    +\left( \frac{(y|y)}{(y|b)}\right)^2} & < 2~{\rm rad/m},\label{eq:s2_condition_2}
\end{align}
for ${}^{15}{\rm O}$. 
In order to achieve these ion-optical properties, 
first order ion-optical calculations using 
COSY INFINITY~\cite{Makino2006346} were performed. 
In the following subsections,
details of the ion-optical designs are described.

\subsection{Ion-optical system from S0 to S1}
\label{sec:s1focalplane}

\begin{figure}[t]
\centering
\includegraphics[width=0.85\linewidth]{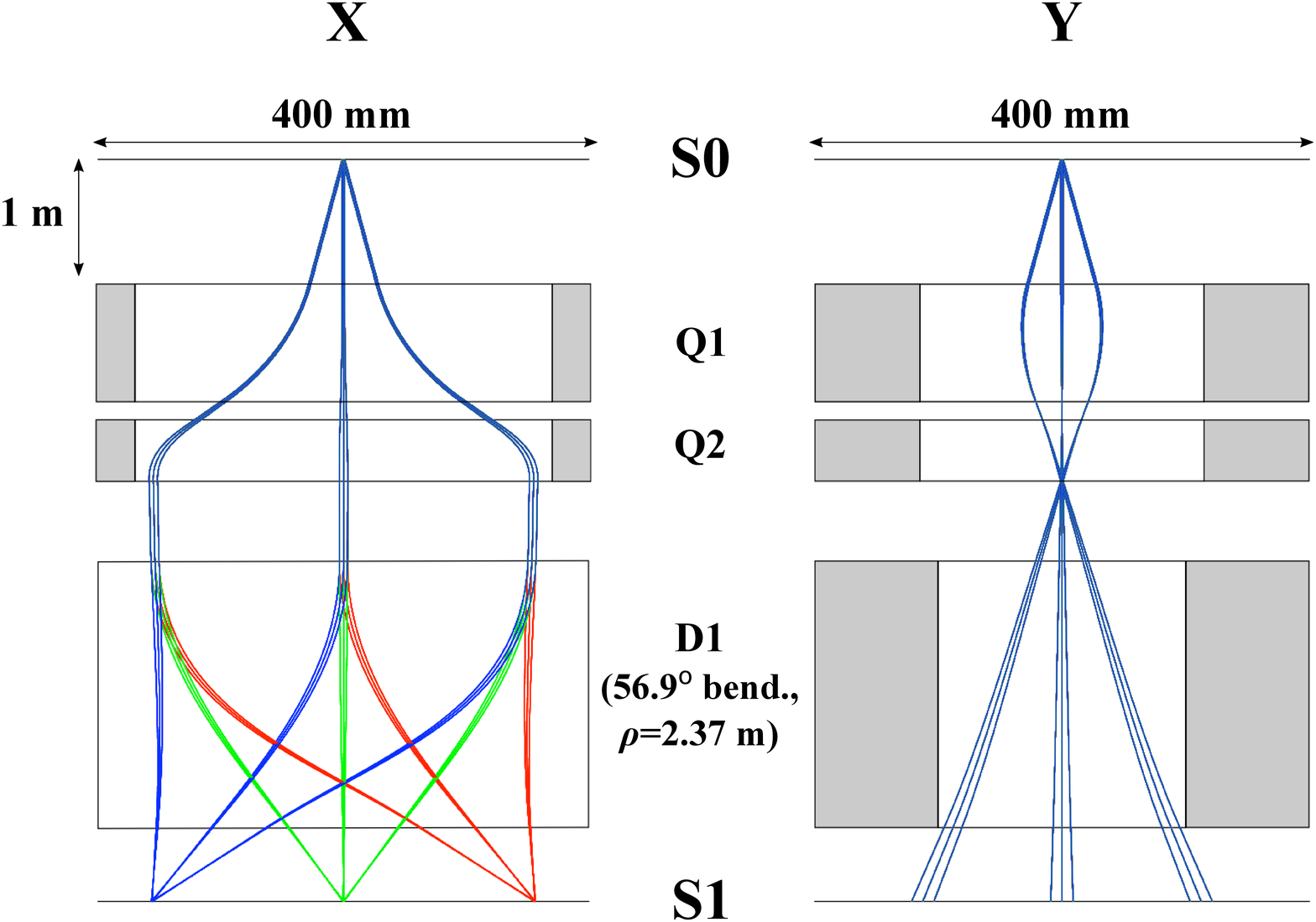}
\caption{\label{fig:s1trajectory} 
  Results of the ion-optical calculations 
  for the particle trajectories from S0 to S1.  
  Left and right panels represent 
  horizontal and vertical trajectories, respectively, 
  for the particles with 
  $x_{\rm S0} = \{ 0, \pm 1~{\rm mm} \}$, 
  $y_{\rm S0} = \{ 0, \pm 1~{\rm mm} \}$, 
  $a_{\rm S0} = \{ 0, \pm 25~{\rm mrad} \}$, 
  $b_{\rm S0} = \{ 0, \pm 25~{\rm mrad} \}$, 
  and 
  $\delta = \{ 0, \pm 10\% \}$.
  }
\end{figure}

In order to optimize the ion-optical properties 
for the particle trajectories from S0 to S1, 
the ion-optical calculations were performed 
in the Q1-Q2-D1 magnet configuration. 
In the calculations, 
the D1 magnet, 
which is a $32.7^{\circ}$ bending magnet with a radius of $\rho=4.4~{\rm m}$ 
in the standard operation of SHARAQ, 
was used as a $56.9^{\circ}$ bending magnet 
with a radius of $\rho = 2.37~{\rm m}$. 
The results of the first order ion-optical calculations 
are shown in Fig.~\ref{fig:s1trajectory}.
In the left and right panels, 
the horizontal and vertical trajectories are shown 
for the particles with 
$x_{\rm S0} = \{ 0, \pm 1~{\rm mm} \}$, 
$y_{\rm S0} = \{ 0, \pm 1~{\rm mm} \}$, 
$a_{\rm S0} = \{ 0, \pm 25~{\rm mrad} \}$, 
$b_{\rm S0} = \{ 0, \pm 25~{\rm mm} \}$, 
and 
$\delta = \{ 0, \pm 10\% \}$. 
Here $x_{\rm S0}$ ($y_{\rm S0}$) and $a_{\rm S0}$ ($b_{\rm S0}$) are 
the horizontal (vertical) position and angle at the target, 
respectively. 
The field strengths of Q1 and Q2 are optimized 
to take a horizontal focus ($(x|a)=0$) and 
satisfy the condition of Eq.~(\ref{eq:s1_condition_2}).
$(y|b)$ has a large value of $-4.5$ 
so that a high angular resolution is achieved in the vertical direction,
as can be seen in the right panel of Fig.~\ref{fig:s1trajectory}. 

Table~\ref{tab:s1prm} summarizes 
the ion-optical properties for the particle trajectories from S0 to S1. 
The ratio of the dispersion [$(x|\delta)_{\rm S1}=-1.56~{\rm m}$] and 
the horizontal magnification [$(x|x)_{\rm S1} = -0.36$] is 
$(x|\delta)_{\rm S1}/(x|x)_{\rm S1} =  4.33~{\rm m}$, 
which satisfies the condition of Eq.~(\ref{eq:s1_condition_1}). 
The resulting resolving power is $p/\delta p = 4330$ 
for a monochromatic image size of $\Delta x_{\rm S0} = 1~{\rm mm}$. 
Because of a large value of $|(a|a)_{\rm S1}| = 2.75$, 
a high angular resolution of $\Delta a < 1~{\rm mrad}$ 
can be achieved in the horizontal direction. 
On the other hand, in the vertical direction, 
the angular resolution is somewhat worse 
($\Delta b = 2~{\rm mrad}$) 
due to a large value of $|(y|y)_{\rm S1}| = 9.00$.

In the actual trajectories from S0 to S1,  
the particles pass 
far from the central orbit of the D1 magnet, 
where a field homogeneity is not good. 
Therefore, we should consider the effects of higher order aberrations. 
For this purpose, 
the measurements using a proton beam were performed. 
The details of the measurements will be described in 
Section~\ref{sec:ionopticsstudy}. 

\begin{table*}[t]
  \small
\begin{center}
  \begin{tabular}{lc}
\hline
\hline
Operation mode & separated flow mode \\
\hline
$(x|x)_{\rm S1}$              & -0.36 \\
$(x|a)_{\rm S1}$ [m/rad]      &  0.00 \\
$(x|\delta)_{\rm S1}$ [m]     & -1.56 \\
$(a|x)_{\rm S1}$ [rad/m]      & -1.53 \\
$(a|a)_{\rm S1}$              & -2.75 \\
$(a|\delta)_{\rm S1}$ [rad]   & -0.75 \\
$(y|y)_{\rm S1}$              & -9.00 \\
$(y|b)_{\rm S1}$ [m/rad]      & -4.50 \\
Resolving power (for image size of 1~mm)
                             & 4330  \\
Horizontal angular resolution (for image size of 1~mm) [mrad]
                             &  $< 1$ \\
Vertical angular resolution (for image size of 1~mm) [mrad]
                             &  2  \\
Momentum acceptance          & $\pm 12\%$ \\
Horizontal acceptance [mrad] & $\pm 27$ \\
Vertical acceptance [mrad]   & $\pm 25$ \\
Solid angle [msr]            &  2.2  \\
\hline
\hline
\end{tabular}
\end{center}
\caption{Design parameters of the ion-optical system from S0 to S1.}
  \label{tab:s1prm}
\end{table*}

\subsection{Ion-optical system from S0 to S2}
\label{sec:s2focalplane}

\begin{figure}[t]
\centering
\includegraphics[width=0.85\linewidth]{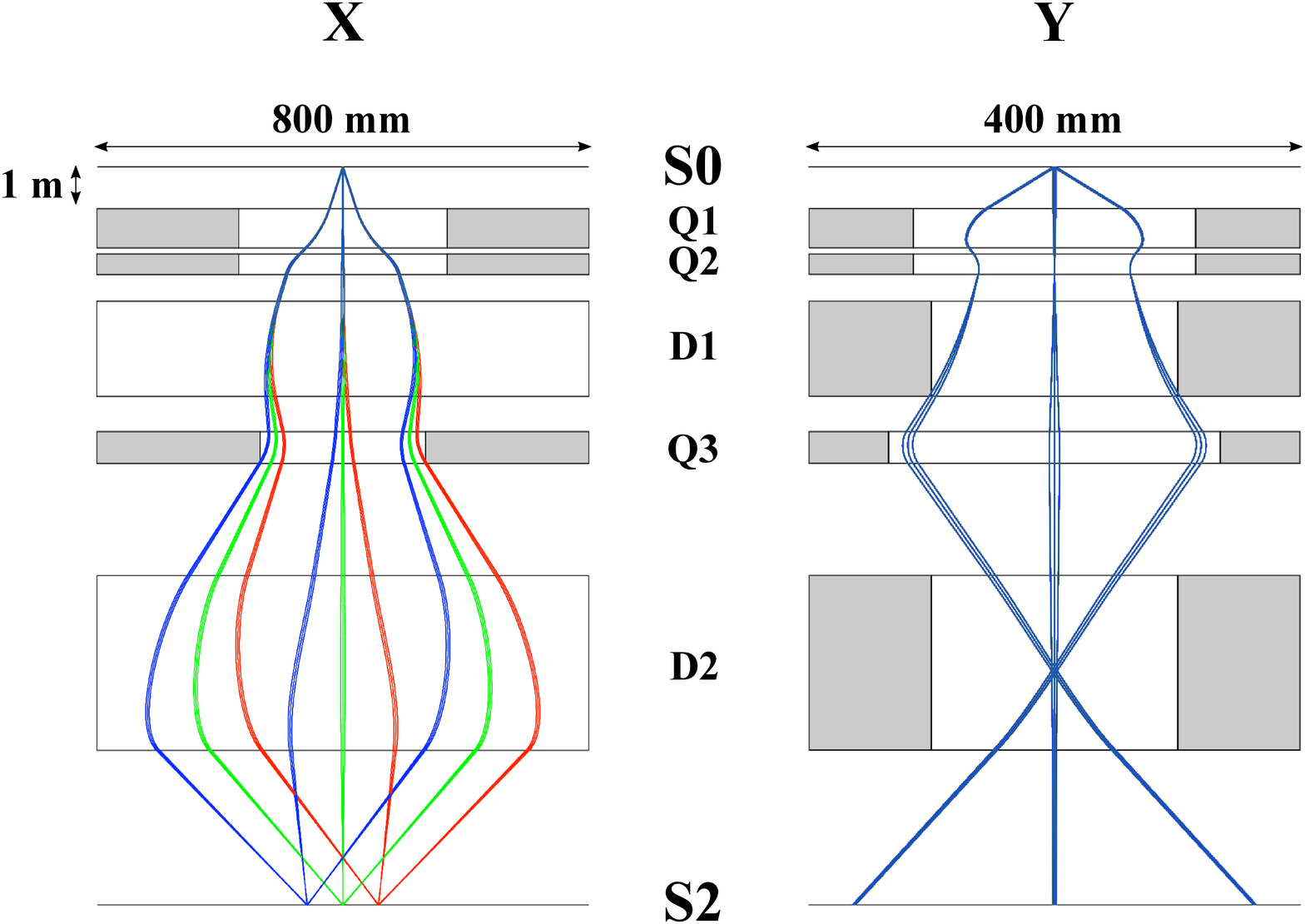}
\caption{\label{fig:s2trajectory}
Results of the ion-optical calculations for the particle trajectories from 
S0 to S2. 
Left and right panels represent 
horizontal and vertical trajectories, respectively, 
for the particles with 
$x_{\rm S0} = \{ 0, \pm 1~{\rm mm} \}$,
$y_{\rm S0} = \{ 0, \pm 1~{\rm mm} \}$,
$a_{\rm S0} = \{ 0, \pm 20~{\rm mrad} \}$,
$b_{\rm S0} = \{ 0, \pm 50~{\rm mm} \}$,
and 
$\delta = \{ 0, \pm 1\% \}$.
}
\end{figure}

The results of the first order ion-optical calculations for 
the particle trajectories from S0 to S2
are shown in Fig.~\ref{fig:s2trajectory}. 
Left and right panels represent 
the horizontal and vertical trajectories 
for the particles with 
$x_{\rm S0} =  \{ 0, \pm 1~{\rm mm} \}$, 
$y_{\rm S0} =  \{ 0, \pm 1~{\rm mm} \}$, 
$a_{\rm S0} =  \{ 0, \pm 20~{\rm mrad} \}$, 
$b_{\rm S0} =  \{ 0, \pm 50~{\rm mm} \}$, 
and 
$\delta = \{ 0, \pm 1\% \}$.
It should be noted that 
the field strengths of Q1 and Q2 are optimized
to satisfy the ion-optical design for the trajectories from S0 to S1. 
The remaining parameter Q3 is adjusted to take a horizontal focus
[$(x|a)_{\rm S2}=0$] at the S2 focal plane. 
The ion-optical properties are summarized in Table~\ref{tab:s2prm}.
We can see that the designed parameters satisfy 
the conditions of Eqs.~(\ref{eq:s2_condition_1}) and (\ref{eq:s2_condition_2}). 
For comparison, the design parameters for the standard mode are also shown. 
The new ion-optical design keeps a high resolving power 
of $p/\delta p = 15300$ and 
a high angular resolution of $< 1~{\rm mrad}$,
while the horizontal angular acceptance is about 30\% smaller 
than that for the standard mode. 
We also performed the ion-optical calculations 
in combination with 
the SHARAQ high-resolution beam line~\cite{Kawabata20084201}, 
and found that 
the lateral and angular dispersion-matching conditions are satisfied
also for the new operation mode.

\begin{table*}[t]
  \small
\begin{center}
  \begin{tabular}{lcc}
\hline
\hline
Operation mode & separated flow mode & standard mode \\
\hline
$(x|x)_{\rm S2}$              & -0.38 & -0.40 \\
$(x|a)_{\rm S2}$ [m/rad]      &  0.00 &  0.00 \\
$(x|\delta)_{\rm S2}$ [m]     & -5.81 & -5.86 \\
$(a|x)_{\rm S2}$ [rad/m]      & -0.74 & -0.87 \\
$(a|a)_{\rm S2}$              & -2.65 & -2.50 \\
$(a|\delta)_{\rm S2}$ [rad]   &  0.65 &  0.65 \\
$(y|y)_{\rm S2}$              & -1.11 &  0.00 \\
$(y|b)_{\rm S2}$ [m/rad]      & -3.26 & -2.20 \\
Resolving power (for image size of 1~mm)
                             & 15300 & 14700 \\
Angular resolution (for image size of 1~mm) [mrad]
                             &  $< 1 $  & $< 1$ \\
Momentum acceptance          & $\pm 1\%$ & $\pm 1\%$ \\
Horizontal acceptance [mrad] & $\pm 20$ & $\pm 30$\\
Vertical acceptance [mrad]   & $\pm 50$ & $\pm 50$\\
Solid angle [msr]            &  3.2 & 4.8 \\
\hline
\hline
\end{tabular}
\end{center}
\caption{Design parameters of the ion-optical system from S0 to S2. 
  The design parameters for the standard mode of SHARAQ are also shown 
  for comparison. 
}
\label{tab:s2prm}
\end{table*}

\subsection{Dispersion mismatch between
  beam line and ion-optical system from S0 to S1}
When the disperion-matched transport is used with the separated flow mode, 
the lateral and angular dispersion-matching conditions are satisfied 
between the beam line and the ion-optical system from S0 to S2. 
These conditions are exprressed as 
\begin{align}
  (x|x)_{\rm S2} (x|\delta)_{\rm bl}
  + (x|a)_{\rm S2} (a|\delta)_{\rm bl}
  + (x|\delta)_{\rm S2}   =  0, \label{eq:dmcondition1}\\
  (a|x)_{\rm S2} (x|\delta)_{\rm bl}
  + (a|a)_{\rm S2} (a|\delta)_{\rm bl}
  + (a|\delta)_{\rm S2}   =  0, \label{eq:dmcondition2}
\end{align}
where $(x|\delta)_{\rm bl}$ and $(a|\delta)_{\rm bl}$
are the matrix elements in the beam line. 
Then $(x|\delta)_{\rm bl}=-15.3~{\rm m}$ and $(a|\delta)_{\rm bl}=4.5~{\rm rad}$
are obtained from Eqs.~(\ref{eq:dmcondition1}) and (\ref{eq:dmcondition2}).
Apparently, the dispersion-matching conditions can not be satisfied at S1
with the same beam-line setting as: 
\begin{align}
  (x|x)_{\rm S1} (x|\delta)_{\rm bl}
  + (x|a)_{\rm S1} (a|\delta)_{\rm bl}
  + (x|\delta)_{\rm S1}   =  3.9~{\rm m}, \label{eq:dmmismatch1}\\
  (a|x)_{\rm S1} (x|\delta)_{\rm bl}
  + (a|a)_{\rm S1} (a|\delta)_{\rm bl}
  + (a|\delta)_{\rm S1}   =  10.2~{\rm rad} \label{eq:dmmismatch2}. 
\end{align} 
The non-zero values of Eqs.~(\ref{eq:dmmismatch1}) and (\ref{eq:dmmismatch2}) 
bring uncertainties in $\delta$ and $a_{\rm S0}$ of protons, respectively, 
according to the momentum spread of the beam $\Delta p/p$. 
Since $\Delta p/p$ is typically $0.1\%$, 
the ambiguities are estimated to be 
$\Delta x_{\rm S1} = 3.9~{\rm mm}$ and $\Delta a_{\rm S1}=10.2~{\rm mrad}$ at S1,
which correspond to
the ambiguities of $\Delta \delta=1/400$ and $\Delta a_{\rm S0}=3.7~{\rm mrad}$ 
in the reconstruction of Eqs.~(\ref{eq:delta_reconst}) and (\ref{eq:a_reconst}). 
These ambiguities are comparable to our required resolutions. 
Therefore we conclude that the dispersion mismatch between 
the beam line and the ion-optical system from S0 to S1 
has no significant effect on our requirements. 

\section{Experiments}
\label{sec:experiments}

\subsection{Experimental setup at S1}
\label{sec:experimental_setup_at_s1}
For the measurements with the separated flow mode of SHARAQ, 
we have developed a tracking detector system at the S1 focal plane. 
This system consists of two multi-wire drift chambers (MWDCs) 
and two plastic scintillators, as shown in Fig.~\ref{fig:setup}.
Table~\ref{tab:mwdcspec} shows the specifications of the MWDCs. 
This MWDC is basically an atmospheric operational version 
of low-pressure multi-wire drift chambers (LP-MWDCs) 
in Ref.~\cite{Miya2013701}. 
Each MWDC has an X-X'-Y-Y' configuration 
and an effective area of 
$480~{\rm mm}^{W} \times 240~{\rm mm}^{H}$. 
The position resolution and the detection efficiency are
typically $300~\mu{\rm m}$ (FWHM) and $>95\%$, respectively,
for $250~{\rm MeV}$ protons. 
The MWDCs are installed with a separation of 200~mm
crossing the S1 focal plane. 
Thus the position and angular resolutions in the S1 focal plane 
are typically $200~\mu{\rm m}$ and $2~{\rm mrad}$, respectively. 
The readout electronics and data acquisition (DAQ) system 
are the same as those described in Ref.~\cite{Miya2013701}.
The vaccum and the air are separated 
at the exit of the D1 magnet 
by a kapton film with a thickness of $125~\mu{\rm m}$,
which causes an angular straggling of about 1~{\rm mrad}
for $250~{\rm MeV}$ protons.

\begin{table}[t]
  \small
\begin{center}
\begin{tabular}{ll}
\hline    
\hline 
Configuration & X - X' - Y - Y' \\
Effective area  & $480~{\rm mm}^{W} \times 240~{\rm mm}^{H}$ \\
Cell size  & $12~{\rm mm}^{W} \times 10~{\rm mm}^{t}$ \\
Numbers of channels & 120 \\
Anode wire & Au-W, 20~$\mu {\rm m}^{\phi}$ \\
Potential wire & Cu-W, 80~$\mu {\rm m}^{\phi}$ \\
Cathode plane & Al-Mylar, 2~$\mu {\rm m}^{t}$ \\
Counter gas & P10 : Ar - ${\rm CH}_4$ (90 - 10), 1~atm \\
Gas window & Al-Mylar, 25~$\mu {\rm m}^{t}$ \\
Potential Voltage & $-1.65~{\rm kV}$ \\
Cathode Voltage & $-1.50~{\rm kV}$ \\
\hline
\hline \\
\end{tabular}
\end{center}
\caption{Specifications of the MWDCs at S1. 
The X' (Y') plane is offset by half cell 
from the X (Y) plane.}
\label{tab:mwdcspec}
\end{table}

\subsection{Ion-optics study with proton beam}
\label{sec:ionopticsstudy}

In order to study the ion-optical properties between S0 and S1, 
we measured the trajectories from S0 to S1 by using a proton beam. 
A secondary proton beam at 246~MeV 
was produced by the BigRIPS~\cite{Kubo200397} 
using a primary ${}^{16}{\rm O}$ beam at 
$247~{\rm MeV/{\rm u}}$ 
with a typical intensity of 10~particle~nA 
in a 4-mm thick Be target. 
The secondary beam was transported to the SHARAQ spectrometer 
by using the high-resolution beam line with 
the high-resolution achromatic transport mode~\cite{Michimasa2013305}.  
A typical intensity of the secondary beam was $1~{\rm kcps}$ at S0.
In order to track the beam trajectories, 
LP-MWDCs~\cite{Miya2013701} and
MWDCs described in Sec.~\ref{sec:experimental_setup_at_s1} 
were used at S0 and S1, respectively. 
The beam profile measured at S0 was 
$\sigma_{x}=6~{\rm mm}$ and $\sigma_{a}=11~{\rm mrad}$
in the horizontal direction, and
$\sigma_{y}=8~{\rm mm}$ and $\sigma_{b}=5~{\rm mrad}$ 
in the vertical direction. 
The momentum spread of the beam was about $\pm 0.1\%$. 

\begin{figure}[t]
\centering
\includegraphics[width=0.9\linewidth]{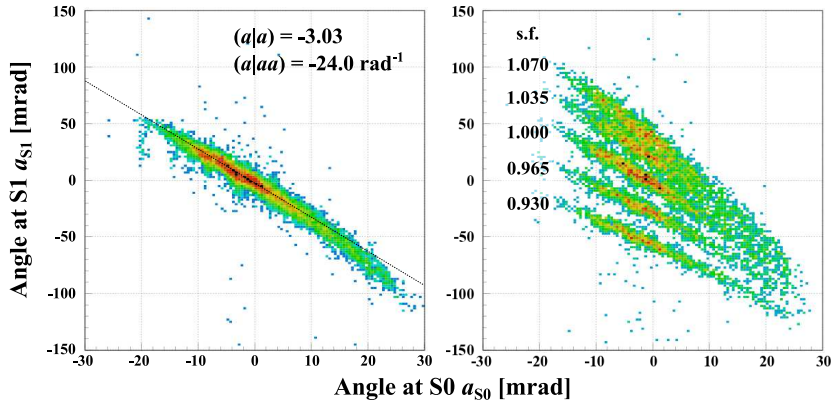}
\caption{\label{fig:aa_correlation}
  Correlation between the horizontal angle at the focal plane S1 and
  the horizontal angle at the focal plane S0 for a proton beam.}
\end{figure}

We obtained the transfer matrix elements from S0 to S1 
by using the correlations of 
the measured beam trajectories at S0 and S1. 
For example, a correlation between 
the horizontal angle at S1 ($a_{\rm S1}$) and
the horizontal angle at S0 ($a_{\rm S0}$) 
provides us with a measure of the $(a|a)$ element. 
The left panel in Fig.~\ref{fig:aa_correlation} shows 
the correlation between $a_{\rm S1}$ and $a_{\rm S0}$.
The gradient of the correlation corresponds to the $(a|a)$ element. 
The second order gradient is also seen,
which corresponds to the $(a|aa)$ element. 
Thus, $(a|a)=-3.03 \pm 0.01$ and $(a|aa)=-24.0 \pm 0.8~{\rm rad}^{-1}$ 
are obtained in this case. 
We also determined the $\delta$ dependent terms of the matrix elements 
by scaling the magnet field of the spectrometer 
because scaling of the magnet field changes $\delta$ effectively. 
The right panel in Fig.~\ref{fig:aa_correlation} shows 
the $a_{\rm S1}$-$a_{\rm S0}$ correlations 
for various magnetic field settings. 
Five loci in the figure represent 
the events for scaling factors of 
0.930, 0.965, 1.000, 1.035, and 1.070, 
which correspond to $\delta = +7.0\%, +3.5\%, 0.0\%, -3.5\%,$ and $-7.0\%$,
respectively. 
The difference of the gradients between the five loci 
corresponds to the matrix elements 
such as $(a|a \delta )$ and $(a|aa \delta )$. 
The matrix elements such as $(a|\delta)$ and $(a|\delta \delta)$ are
also obtained from 
the shift of the $a_{\rm S1}$ positions at $a_{\rm S0}=0~{\rm mrad}$.

Table~\ref{tab:measured_matrix_elements} summarize 
the measured matrix elements from S0 to S1.
For the first order terms, 
the measured matrix elements are in good agreement with 
the design values in Table~\ref{tab:s1prm}. 
We also found that 
the effects of the higher order terms  
were enough large to change the trajectories, 
and that these effects should be correctly taken into account. 

\begin{figure}[t]
\centering
\includegraphics[width=0.7\linewidth]{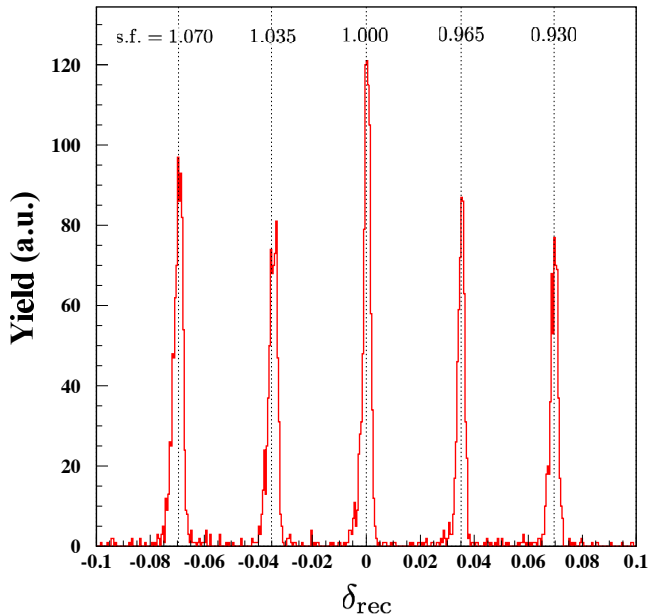}
\caption{\label{fig:proton_reconstructed_delta}
  The $\delta$ values reconstructed from the beam trajectory at S1.}
\end{figure}

We reconstructed the beam momentum and angles at S0 
from the beam trajectory at S1 
using an inverse matrix determined 
by the measured matrix elements. 
In the reconstruction, the events 
in a region of $|x_{\rm S0}| < 0.5~{\rm mm}$ and 
$|y_{\rm S0}| <0.5~{\rm mm}$ were selected, 
and $x_{\rm S0}$ and $y_{\rm S0}$ values were  
set to $x_{\rm S0}=y_{\rm S0}=0$. 
Figure~\ref{fig:proton_reconstructed_delta} shows 
the $\delta$ values reconstructed from the beam trajectory at S1. 
Five peaks correspond to the events for different magnet field settings. 
The $\delta$ value for each peak is well reproduced 
according to the corresponding scaling factor.
The width for each peak varies between 
$0.28\%$ and $0.33\%$ FWHM depending on $\delta$,
which is mainly due to the momentum spread of the beam of about $\pm 0.1\%$. 

\begin{figure}[t]
\centering
\includegraphics[width=0.85\linewidth]{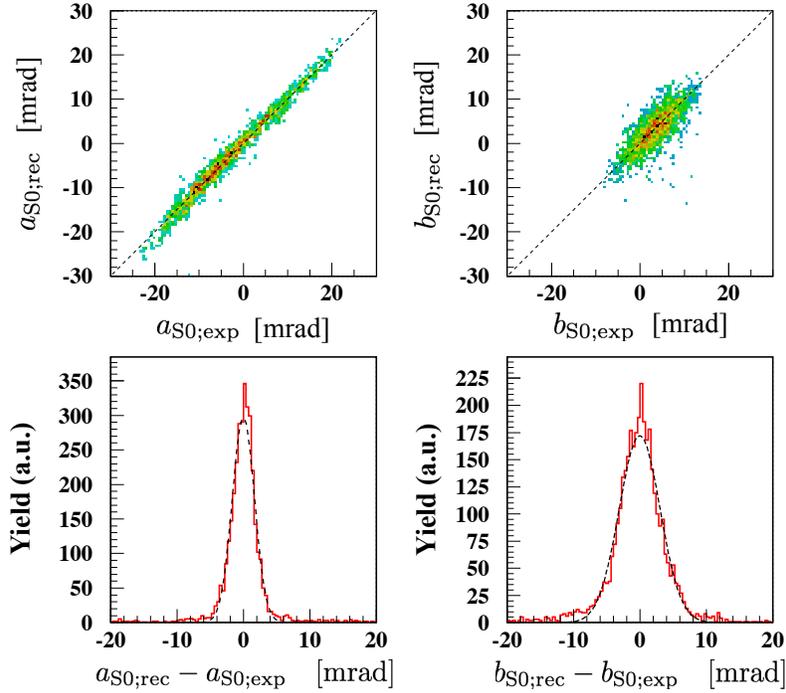}
\caption{\label{fig:proton_reconstructed_angle}
  The $a_{\rm S0}$ and $b_{\rm S0}$ values reconstructed
  from the beam trajectory at S1. 
  Top panels show the correlations
  between the reconstructed and measured values,
  and bottom panels take the differences between these values. 
}
\end{figure}

We also compared 
the $a_{\rm S0}$ and $b_{\rm S0}$ values
reconstructed from the beam trajectory at S1 
with those measured by MWDCs at S0 
in Fig.~\ref{fig:proton_reconstructed_angle}. 
Top panels show the correlations between 
the reconstructed and measured values, 
and bottom panels take the differences between these values. 
We can see that 
both the $a_{\rm S0}$ and $b_{\rm S0}$ values are properly reconstructed. 
The widths of the difference distributions are 
3.8 and 7.2~mrad FWHM 
for $a_{\rm S0}$ and $b_{\rm S0}$, respectively, 
main parts of which arise from
the position and angular resolutions at S0
(3~mm and 2~mrad FWHM).

\begin{table*}[t]
  \small
  \begin{center}
    \begin{tabular}{lclclc}
  \hline \hline
  \multicolumn{2}{c}{$x$} &
  \multicolumn{2}{c}{$a$} & 
  \multicolumn{2}{c}{$y$} 
  \\
  \hline
  $(x|x)_{\rm S1}$     & $-0.34 \pm 0.01$ 
  & 
  $(a|x)_{\rm S1}$     & $-1.43 \pm 0.01$
  & 
  $(y|y)_{\rm S1}$     & $-9.55 \pm 0.02$
  \\
  $(x|a)_{\rm S1}$     &  $0.01 \pm 0.01$
  & 
  $(a|a)_{\rm S1}$     & $-3.03 \pm 0.01$
  & 
  $(y|b)_{\rm S1}$     & $-4.70 \pm 0.05$ 
  \\
  $(x|\delta)_{\rm S1}$     & $-1.5703 \pm 0.0002$
  & 
  $(a|\delta)_{\rm S1}$     & $-0.70 \pm 0.05$
  &  &
  \\
  $(x|aa)_{\rm S1}$     & $0.80 \pm 0.74$ 
  & 
  $(a|aa)_{\rm S1}$     & $-24.0 \pm 0.8$ 
  & 
  $(y|ab)_{\rm S1}$     & $-36 \pm 3$ 
  \\
  $(x|a \delta)_{\rm S1}$     & $0.40 \pm 0.14$
  &
  $(a|a \delta)_{\rm S1}$     & $11.5 \pm 0.2$
  & 
  $(y|y \delta)_{\rm S1}$     & $34.0 \pm 0.4$
  \\
  $(x|\delta \delta)_{\rm S1}$     & $-7.319 \pm 0.001$
  &
  $(a|\delta \delta)_{\rm S1}$     & $1.5 \pm 0.2$
  & 
  $(y|b \delta)_{\rm S1}$     & $23.5 \pm 0.9$
  \\
  $(x|aaa)_{\rm S1}$     & $-820 \pm 31$
  &
  $(a|aa \delta)_{\rm S1}$     & $80 \pm 16$
  & 
  $(y|ab \delta)_{\rm S1}$     & $231 \pm 73$
  \\
  $(x|a \delta \delta)_{\rm S1}$     & $-57 \pm 1$
  &
  $(a|a \delta \delta)_{\rm S1}$     & $-12 \pm 4$
  & 
  $(y|b \delta \delta)_{\rm S1}$     & $-74 \pm 19$
  \\
  $(x|\delta \delta \delta)_{\rm S1}$     & $-29.23 \pm 0.05$
  &
  $(a|\delta \delta \delta)_{\rm S1}$     & $8 \pm 6$
  & 
  \\
  \hline \hline\\
    \end{tabular}
  \end{center}
  \caption{Measured transfer matrix elements from S0 to S1.
    Units for lengths and angles are in meter and radian, respectively.}
  \label{tab:measured_matrix_elements}
\end{table*}


\subsection{Measurement of $({}^{16}{\rm O},{}^{16}{\rm F})$ reaction}
\label{sec:paritytransfer_experiment}

The performance of the separated flow mode
was studied by using the $({}^{16}{\rm O},{}^{16}{\rm F})$ reaction. 
A primary ${}^{16}{\rm O}$ beam at $247~{\rm MeV/u}$ 
and $10^7~{\rm pps}$ was transported to the S0 target position. 
The beam line to the spectrometer was set up 
for the dispersion matched transport. 
A plastic scintillator with a thickness of $1~{\rm mm}$ 
was used as a reaction target. 
The outgoing ${}^{15}{\rm O}+p$ produced by the decay of ${}^{16}{\rm F}$ 
were measured in coincidence. 
The particles were momentum analyzed by using the SHARAQ spectrometer. 
The ${}^{15}{\rm O}$ particles were detected with two LP-MWDCs 
at the S2 focal plane, 
while the protons were detected with two MWDCs at the S1 focal plane. 

\begin{figure}[t]
\centering
\includegraphics[width=0.9\linewidth]{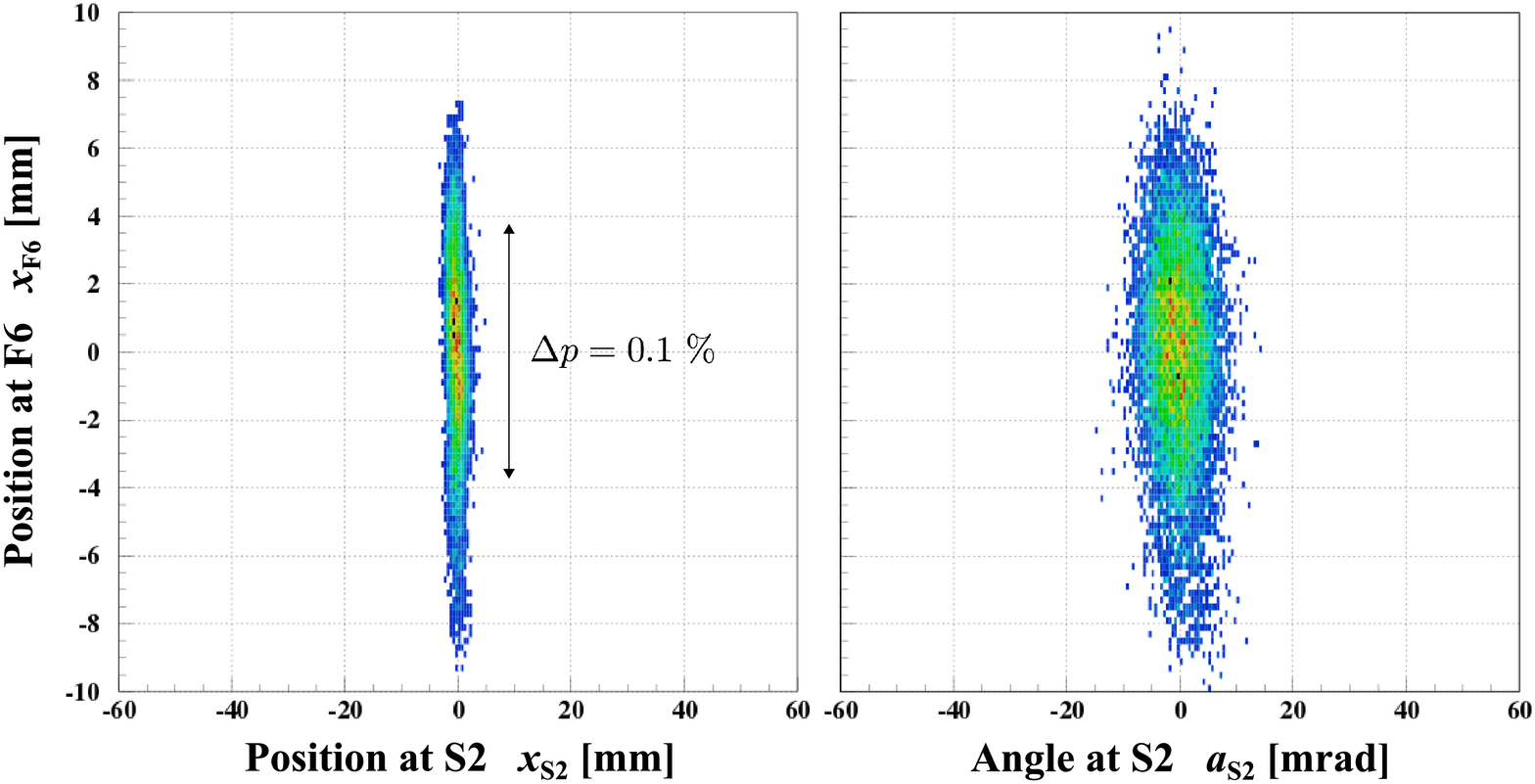}
\caption{\label{fig:dispersion_matching}
  Correlation between $x_{\rm F6}$ and the position (left) and angle (right)
  at S2 for a ${}^{16}{\rm O}$ beam at 247~{\rm MeV/u}. 
  Upright correlations observed in the figures indicate that
  the lateral and angular dispersion-matching conditions
  are fulfilled.}
\end{figure}

Figure~\ref{fig:dispersion_matching} represents the correlations 
between the position (angle) at S2 and the position at F6 
for the ${}^{16}{\rm O}$ beam. 
Here F6 is a dispersive focal plane of the beam line, 
and the position was measured by 
a parallel plate avalanche counter (PPAC). 
The position at F6 corresponds to the beam momentum. 
Upright correltions found in the figures indicate that 
the position and angle at S2 are independent of the beam momentum, 
which demonstrates that 
the lateral and angular dispersion-matching conditions
are satisfied also for the separated flow mode.

\begin{figure}[t]
\centering
\includegraphics[width=0.7\linewidth]{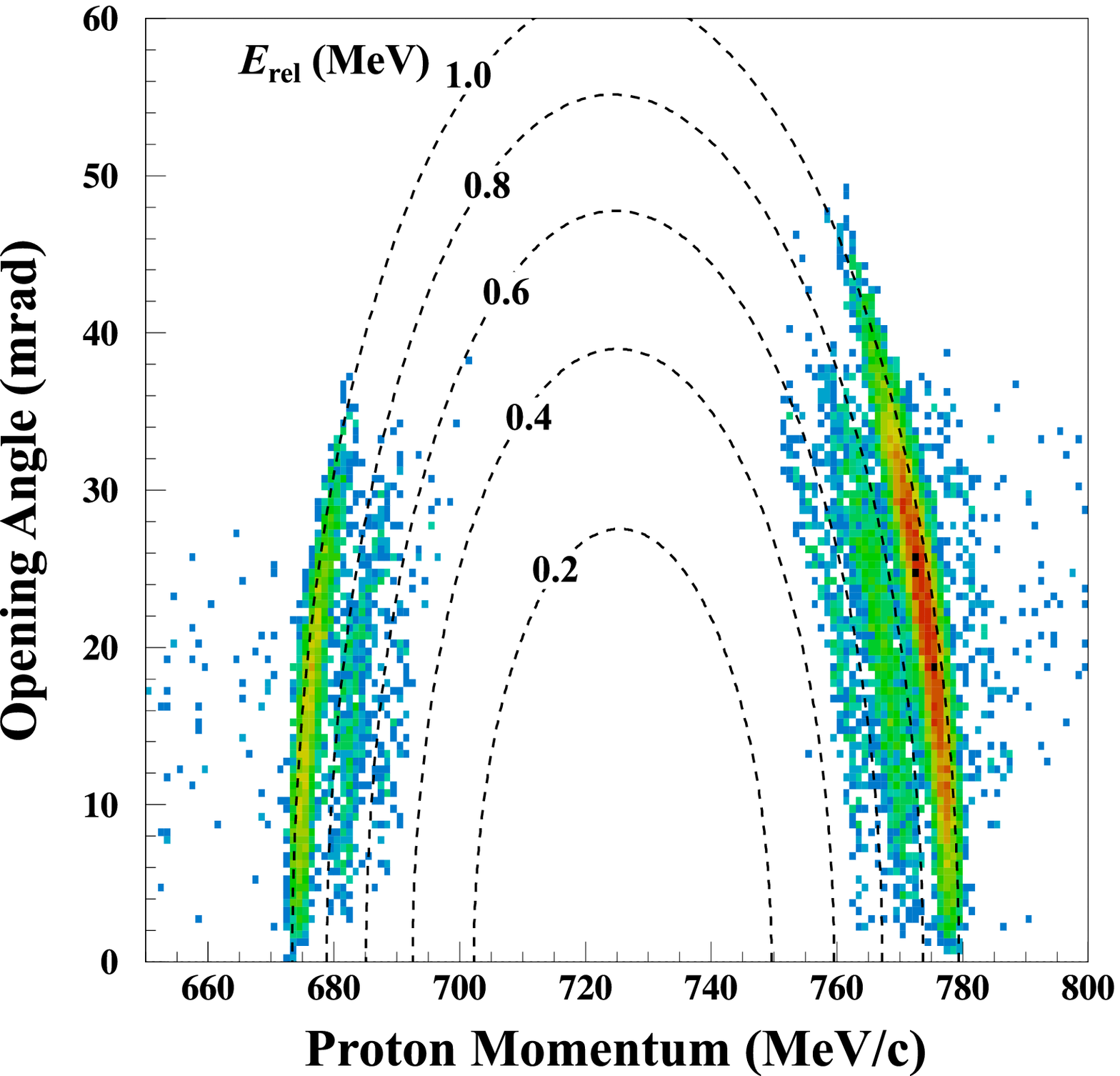}
\caption{\label{fig:pmom_openang}
Two-dimentional plot of 
the proton momentum and 
the opening angle between the $^{15}{\rm O}$ particle and the proton. 
The dashed curves indicate the relative energy between 
the ${}^{15}{\rm O}$ particle and the proton in $0.2~{\rm MeV}$ step. 
}
\end{figure}

Figure~\ref{fig:pmom_openang} shows 
a two-dimentional plot of 
the proton momentum and 
the opening angle between the $^{15}{\rm O}$ particle and the proton. 
The dashed curves indicate the relative energy $E_{\rm rel}$ between 
the ${}^{15}{\rm O}$ particle and the proton in $0.2~{\rm MeV}$ step. 
As can be seen, 
three states of ${}^{16}{\rm F}$ with
$J^{\pi}~=~0^-, 1^-$, and $2^-$ 
at $E_{\rm rel}~=~0.535, 0.728$, and 0.959~{\rm MeV} 
are clearly separated. 
The obtained $E_{\rm rel}$ resolution is 
100~keV FWHM at $E_{\rm rel}~=~0.535~{\rm MeV}$, 
which satisfies our requirement.

\begin{figure}[t]
\centering
\includegraphics[width=0.7\linewidth]{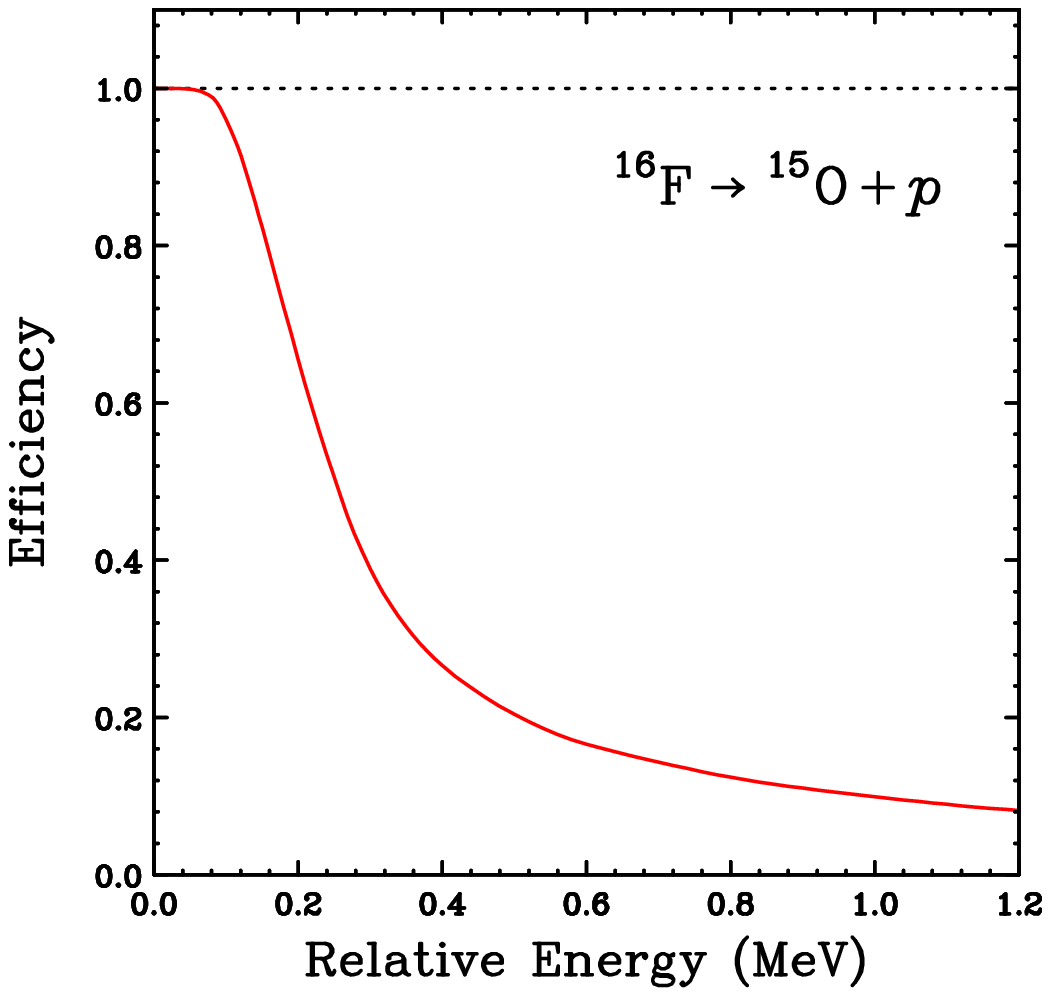}
\caption{\label{fig:acpeff}
  Detection efficiency for the $^{15}{\rm O}+p$ coincidence events 
  calculated by Monte Carlo simulations,
  as a funciton of relative energy. 
}
\end{figure}

We also estimated the detection efficiency 
for the $^{15}{\rm O}+p$ coincidence events 
from the Monte Carlo simulation, 
where the acceptance and the finite resolution in angles and momenta
are taken into account. 
The result is shown in Fig.~\ref{fig:acpeff},
as a funciton of $E_{\rm rel}$. 
The obtained efficiency is 0.189 at $E_{\rm rel}=0.535~{\rm MeV}$,
which is mainly due to the angular acceptance for the proton.

\begin{figure}[t]
\centering
\includegraphics[width=0.7\linewidth]{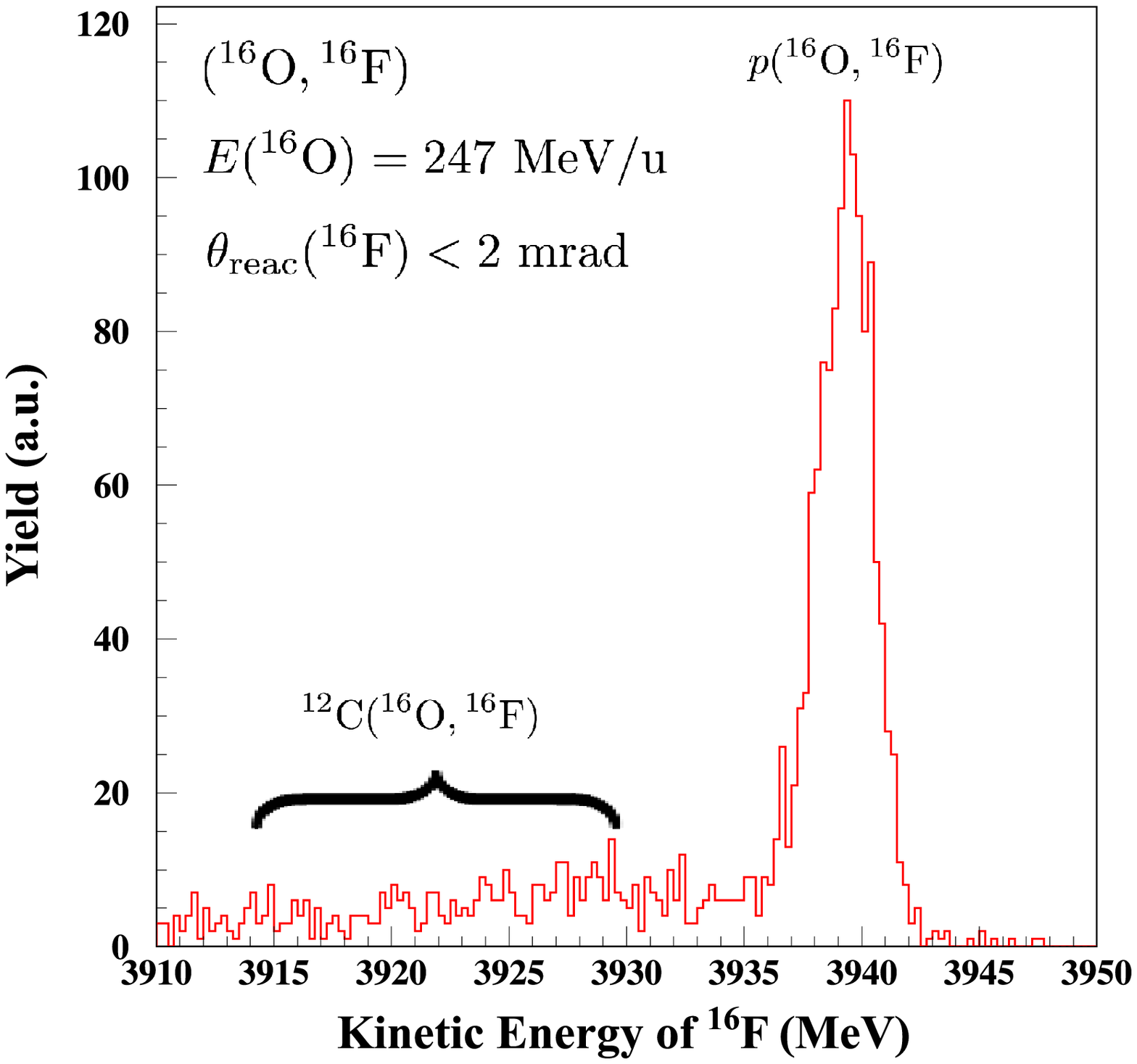}
\caption{\label{fig:16fke}
  Kinetic-energy distribution of ${}^{16}{\rm F}$
  for the $({}^{16}{\rm O},{}^{16}{\rm F})$ reaction at 
  $\theta_{\rm reac}<2~{\rm mrad}$. 
}
\end{figure}

Figure~\ref{fig:16fke} shows 
the kinetic-energy distribution of ${}^{16}{\rm F}$ 
for the $({}^{16}{\rm O},{}^{16}{\rm F})$ reaction 
at a reaction angle of $\theta_{\rm reac} < 2~{\rm mrad}$. 
A prominent peak at $E({}^{16}{\rm F}) \sim 3940~{\rm MeV}$ 
corresponds to the $({}^{16}{\rm O},{}^{16}{\rm F})$ reaction 
on hydrogens in the plastic scintillator. 
The width is about 2.8~MeV FWHM. 
The intrinsic energy resolution is about 2.1~MeV FWHM 
by considering the energy straggling of the particles in the target 
($\sim 1.8~{\rm MeV}$). 
Thus our requirement is satisfied also 
for the ${}^{16}{\rm F}$ energy resolution. 

\section{Summary}
  New operation mode, ``{\it separated flow mode}'', has been developed 
  for in-flight proton decay experiments with the SHARAQ spectrometer. 
  The concept of the separated flow mode is 
  the use of the SHARAQ spectrometer
  as two spectrometers with different magnet configurations,
  which allows the coincidence measurements of
  the proton and heavy-ion pairs produced
  from the decays of proton-unbound states in nuclei. 

  The ion-optical properties of the new mode were studied 
  by using a proton beam at 246~MeV. 
  The transfer matrix elements were experimentally determined 
  including the higher order terms,
  and the momentum vector was properly reconstructed 
  from the parameters measured in the focal plane of SHARAQ. 

  The separated flow mode was successfully introduced 
  in the experiment with the $({}^{16}{\rm O},{}^{16}{\rm F})$ reaction 
  at a beam energy of 247~MeV/u. 
  The outgoing ${}^{15}{\rm O}+p$ produced by the decay of ${}^{16}{\rm F}$ 
  were measured in coincidence with SHARAQ.
  High energy resolutions of 100~keV (FWHM) and $\sim 2~{\rm MeV}$
  were achieved for the relative energy of 535~keV, 
  and the ${}^{16}{\rm F}$ energy of 3940~MeV, respectively.
  Such an accurate missing-mass measurement 
  combined with an invariant-mass method 
  gives a unique opportunity to explore little-studied excitation modes in nuclei 
  by using new types of reaction probes with particle decay channels.

\section*{Acknowledgments}
We thank the technical staff of the accelerator 
and the BigRIPS spectrometer at the RIKEN Nishina Center,
and the accelerator staff at the CNS, the University of Tokyo,
for providing us the excellent beam. 
We also thank K.~Yako for valuable discussions.
This work was supported by JSPS KAKENHI Grant Nos. 23840053 and 14J09731. 


\bibliography{separated_flow.bib}

\end{document}